\documentclass[twocolumn,showpacs,preprintnumbers,amsmath,amssymb,prd,letterpaper,floatfix,nofootinbib,superscriptaddress]{revtex4-1}  
 \usepackage{graphicx}
\usepackage{epsfig}
\usepackage{color}
\usepackage{longtable}
\usepackage{hyperref} 
\usepackage{latexsym}
\usepackage{amsmath}
\usepackage{amssymb}
\usepackage{dsfont}
\usepackage{verbatim}
\usepackage{bm}
 
\begin{document}
\date{\today}

\title{ Study of the  $Z_c^+$ channel using lattice QCD}

\author{Sasa Prelovsek\footnote{sasa.prelovsek@ijs.si}}
\email{sasa.prelovsek@ijs.si}
\affiliation{Department of Physics, University of Ljubljana, Jadranska 19, 1000 Ljubljana, Slovenia}
\affiliation{Jozef Stefan Institute, Jamova 39, 1000 Ljubljana, Slovenia}

\author{C. B. Lang}
\email{christian.lang@uni-graz.at}
\affiliation{Institute of Physics,  University of Graz, A--8010 Graz, Austria}

\author{Luka Leskovec}
\email{luka.leskovec@ijs.si}
\affiliation{Jozef Stefan Institute,  Jamova 39, 1000 Ljubljana, Slovenia}

\author{Daniel Mohler}
\email{dmohler@fnal.gov}
\affiliation{Fermi National Accelerator Laboratory, P.O. Box 500, Batavia, Illinois 60510-5011, USA}

\begin{abstract}
  
Recently experimentalists have discovered several  charged charmonium-like hadrons $Z_c^+$ with unconventional quark content  $\bar cc\bar d u$. We perform a search for $Z_c^+$ with mass below $4.2~$GeV in the channel  $I^G(J^{PC})=1^+(1^{+-})$ using  lattice QCD.  The major challenge is presented by the two-meson states $J/\psi\, \pi$, $\psi_{2S}\pi$, $\psi_{1D}\pi$, $D\bar D^*$, $D^*\bar D^*$, $\eta_c\rho$ that are  inevitably present in this channel. The spectrum of eigenstates is extracted using a number of meson-meson  and diquark-antidiquark interpolating fields. For our pion mass of 266~MeV we find all the expected two-meson states but no additional candidate for $Z_c^+$ below $4.2~$GeV. Possible reasons for not seeing an additional  eigenstate related to $Z_c^+$ are discussed. We also illustrate how a simulation incorporating interpolators with a structure resembling low-lying two-mesons states seems to render a $Z_c^+$ candidate, which is however not robust after further two-meson states around $4.2~$GeV are implemented. 
 
 \end{abstract}
 
 \maketitle



\section{Introduction}

Quantum Chromodynamics (QCD) is the fundamental quantum field theory of quarks and gluons. In its strong coupling regime it should explain the masses and other properties of hadrons. Conventional hadrons are composed of either a valence quark $q$ and antiquark $\bar{q}$ (mesons) or three valence quarks (baryons) on top of the sea of quark-antiquark pairs and gluons. One of the most notable and perhaps surprising features until recently was the complete absence of exotic hadrons like $\bar q\bar qqq$ or $\bar q q qqq$.
 
This has changed due to fascinating experimental discoveries over the past seven years. Most of the newly discovered exotic states have unconventional flavor content, likely $\bar c c\bar d u$, and spin and parity quantum numbers $J^P=1^+$. 
The first of these states was the $Z^+(4430)$,  discovered in 2007 by Belle \cite{Choi:2007wga}, remained unconfirmed by BaBar \cite{Aubert:2008aa}, and was recently confirmed by LHCb \cite{Aaij:2014jqa}.
In recent years several similar states have been found by experiment. The $Z_c^+(3900)\to J/\psi\, \pi^+$ was discovered slightly above $D\bar D^*$ threshold   by BESIII \cite{Ablikim:2013mio},  and was confirmed by Belle \cite{Liu:2013dau} as well as using CLEO-c data \cite{Xiao:2013iha}. The spin and parity of $Z_c^+(3900)$ are unclear, and it may correspond to the same state as $Z_c^+(3885)\to (D\bar D^*)^+$ with $J^P=1^+$ \cite{Ablikim:2013xfr}. The pair $Z_c^+(4020)\to h_c\pi^+$ \cite{Ablikim:2013wzq} and $Z_c^+(4025)\to (D^*\bar D^*)^+$ \cite{Ablikim:2013emm}, which may correspond to the same state,  was found by BESIII slightly above $D^*\bar D^*$ threshold. Their spin and parity are unclear and $J^{P}=1^+$ is preferred. Finally, $Z^+(4200)\to J/\psi\, \pi^+$ was reported in 2014 by Belle \cite{Chilikin:2014bkk} favoring $J^P=1^+$. 
 All these states have G-parity $G\!=\!+1$ while their neutral partners have charge conjugation $C\!=\!-1$. Therefore we focus on the channel with $I^G(J^{PC})=1^+(1^{+-})$.

 On the theoretical side, these states have been related to mesonic molecules, diquark-antidiquark structures, hadrocharmonium \cite{Voloshin:2013dpa} or Born-Oppenheimer tetraquarks \cite{Braaten:2013boa}. However the existence of these states within QCD has not yet been settled.  While these states have been addressed theoretically with a  number of phenomenological approaches like quark models, (unitarized) effective field theory and QCD sum rules (for reviews with references see \cite{Brambilla:2014aaa,Brambilla:2010cs,Liu:2013waa}), these approaches are either not based directly on QCD or they depend on parameters (i.e., low energy constants) that are not present in the QCD Lagrangian. The existence of $Z_c^+$ has never been established from first-principle QCD. The problem is a large magnitude of the strong coupling constant $\alpha_s$ at the hadronic energy scale, hence a perturbative expansion is not successful. Lattice QCD represents the only non-perturbative approach that is based directly on QCD, depending only on parameters $m_q$ and $\alpha_s$ that appear in the QCD Lagrangian. 
  
Therefore it is an urgent theoretical task to establish whether QCD supports the presence of an exotic state with quark content $\bar cc\bar du$ using  ab-initio lattice QCD. In a lattice QCD simulation, the states are identified from discrete energy-levels $E_n$ and in principle all eigenstates with the  given quantum number $I^G(J^{PC})=1^+(1^{+-})$ appear. in searching for $Z_c^+$ candidates the whole finite volume spectrum needs to be identified. In particular eigenstates with significant $J/\psi\, \pi$, $\psi_{2S}\pi$, $\psi_{1D}\pi$, $D\bar D^*$, $D^*\bar D^*$, $\eta_c\rho$ components appear which presents a major challenge.
   
The first lattice simulation aimed at $Z_c^+(3900)$ focused on the region below $4~$GeV and found only two-particle states $J/\psi \pi$ and $D\bar D^*$ and no indication for $Z_c^+(3900)$ \cite{Prelovsek:2013xba}. The second simulation studied $D\bar D^*$ scattering near threshold in the same channel and did not yield any indication for $Z_c^+(3900)$ either \cite{Chen:2014afa}.
 
 In the present paper we extend the search for candidates in the channel $I^G(J^{PC})=1^+(1^{+-})$ up to $4.2~$GeV (for our
 setup with $m_\pi=266$ MeV). The task is to determine the discrete spectrum in this channel with the challenging aim to establish all expected energy levels. The main question is  whether there are any extra energy levels in addition to the number of expected two-meson states. An additional energy level near $E\simeq m_{Z_c}$ would be a definite signature for a $Z_c^+$ with an approximate mass $m_{Z_c}$.
  
This paper is organized as follows.  Section II discusses expected two-meson  states below $4.3~$GeV. Section III details how energy levels and overlaps are extracted, while section IV is dedicated to the results. In section V we summarize cautionary remarks and lessons which may be useful for future lattice simulations, and  we conclude in Section VI.   

\section{Two-meson states in lattice QCD}

In lattice QCD   the states are identified from discrete energy-levels $E_n$ and in principle all finite volume eigenstates with given quantum numbers appear. The eigenstate of interest gives an energy level at  $E_n\simeq m_{Z_c}$ if $Z_c$ exists and is not too broad. However, various two-meson states $M_1(\mathbf{p})M_2(-\mathbf{p})$ have the same quantum numbers which presents a major challenge. Individual momenta are discretized due to the periodic boundary conditions in space. If the two mesons do not interact, then $\mathbf{p}= \!\tfrac{2\pi}{L}\mathbf{k}$ with $\mathbf{k}\in \mathbb{N}^3$, and the energies of $M_1(k)M_2(-k)$ states for $a\to 0$ are
\begin{equation}
\label{E}
E^{n.i.}=E_1(k)+E_2(k)\ ,  \ E_{1,2}(k)=\sqrt{m_{1,2}^2+k (\tfrac{2\pi}{L})^2}\;.
\end{equation}
with $k\equiv \mathbf{k}^2$. 
These values are slightly shifted in presence of the interaction. 
In experiment, these states correspond to the two-meson decay products 
with a continuous energy spectrum. In the current study we neglect possible channels with three or more mesons.
 
Our simulation employs dynamical $u$ and $d$ quarks that correspond to the pion mass $m_\pi\simeq 266~$MeV \cite{Mohler:2012na,Lang:2014yfa}.  
The lattice spacing  is $a\!=\!0.1239(13)~$fm. The rather small box $V=16^3\times 32$ with  $L\simeq 2~$fm may lead to sizable finite volume corrections, but it is responsible for a crucial practical advantage. It  makes the $Z_c^+$ search tractable since it reduces the number of $M_1(k)M_2(-k)$ states in the considered energy range, as discussed in Section \ref{sec_V}. 
 
\begin{figure*}[ht!]
\begin{center}
\includegraphics[width=0.65\textwidth,clip]{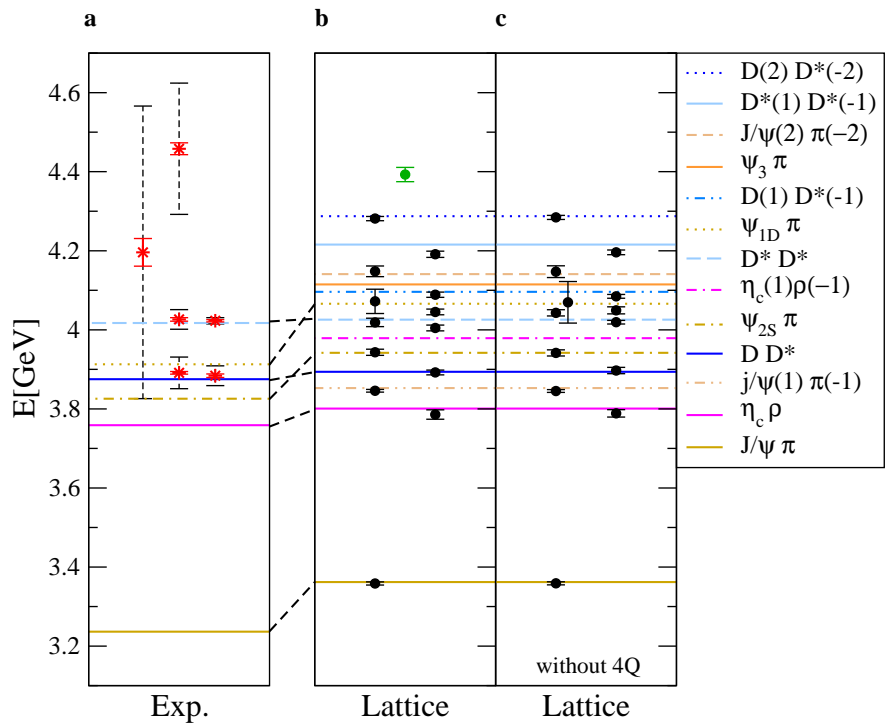} 
\end{center}
\caption{The spectrum for quantum numbers $ I^G(J^{PC})=1^+(1^{+-})$. (a) Position of the experimental $Z_c^+$ candidates \cite{Brambilla:2014aaa}. (b,c)  The discrete energy spectrum from our lattice simulation: (b) shows energies based on complete $22\times 22$  matrix  of interpolators,  (c)  is based on the $18\times 18$ correlator matrix without diquark-antidiquark interpolating fields ${\cal O}^{4q}_{1-4}$ (\ref{O}). The thirteen   lowest lattice energy levels (black circles) are interpreted as two-particle states, which are inevitably  present in a dynamical lattice QCD simulation. No additional   candidate for the exotic $Z_c^+$ is found below $4.2~$GeV.   
The dashed vertical lines indicate twice the experimental widths to illustrate the energy range in which the additional energy level due to $Z_c$ might be expected.}\label{fig:eff}
\end{figure*}

On our lattice (with $m_\pi=266$ MeV) the two-particle states with $I^G(J^P)=1^+(1^+)$ and total momentum zero in the energy region of interest $E\leq 4.3~$GeV are\footnote{We take $E^{lat}_{n}-m_{s.a.}^{lat}+m_{s.a.}^{exp}<4.3~$GeV  as argued below.}
 \begin{align}
 \label{MM}
&J/\psi(0)\pi(0),~\eta_c(0)\rho(0),~J/\psi(1)\pi(-1),~D(0)\bar D^*(0),\nonumber\\
&\psi_{2S}(0)\pi(0), ~D^*(0)\bar D^*(0),\psi_{1D}(0)\pi(0),~\eta_c(1)\rho(-1),\nonumber\\
&D(1)\bar D^*(-1),~\psi_3(0)\pi(0),~J/\psi(2)\pi(-2),~D^*(1)\bar D^*(-1)\nonumber\\
&D(2)\bar D^*(-2)
 \end{align} 
 in order of increasing energy. Their lattice energies $E^{n.i.}$ in the non-interacting limit are denoted by the horizontal lines in Fig. \ref{fig:eff}b and the values follow from the masses and single-meson energies determined on the same   set of gauge configurations \cite{Mohler:2012na,Lang:2011mn}. Establishing two-meson states up to $4.3~$GeV at $m_\pi=266$ MeV should suffice for searching  fairly narrow exotic candidates with mass below $4.07~$GeV for physical pion mass\footnote{The value 4.07GeV results from taking into account a possible pion mass dependence which we estimate conservatively by the behavior of the threshold most sensitive to the pion mass.}.
 
The $\psi_{1D}$ in (\ref{MM}) denotes $\psi(3770)$. The appearance of $\psi_{3} \pi$, where $\psi_3$ denotes the charmonium with $J^{PC}\!=\!3^{--}$, is an artifact due to reduced symmetry on the cubic lattice  as discussed in Appendix \ref{app_a}.   The $h_c(0)\pi(0)$ is not present for $J^P=1^+$ since non-vanishing relative momentum $p$ is required by the orbital momentum $l=1$. The  $h_c(1)\pi(-1)$  lies near $4.25~$GeV, but is not listed in (\ref{MM}) since this is the only two-meson state below $4.3~$GeV that we do not aim to extract due to the arguments given in Appendix \ref{app_a}. The energy of $\rho(-1)$ is extracted from the diagonal correlator of $\bar d \gamma_j u$ neglecting the resonance nature of the $\rho$.
  
Our aim is to extract and identify all two-particle energy-levels (\ref{MM})  from the full, coupled correlator matrix of hadron operators and establish whether QCD predicts additional states related to the exotic $Z_c^+$ hadron. 

This goal presents a considerable challenge by itself. Note that a rigorous treatment (via a L\"uscher-type  finite volume formalism \cite{Luscher:1986pf,Luscher:1990ux,Doring:2011vk,Hansen:2012tf,Briceno:2014oea}) would require the determination of the scattering matrix for all two-particle channels that couple, and a subsequent determination of the mass and the width for any $Z_c^+$ resonance(s). The elastic scattering within a single channel has been rigorously treated by a number of lattice simulations recently. The first lattice simulation aimed at determining scattering matrix for  two-coupled channels \cite{Dudek:2014qha} also shows promise in this respect,  while the rigorous treatment of seven coupled channels is still beyond the capabilities of any lattice simulation at present. 

Therefore we take a simplified approach where the existence of $Z_c^+$ is investigated by analyzing the number of energy levels, their positions and overlaps with the considered lattice operators $\langle \Omega |{\cal O}_j|n\rangle$. The formalism does predict an appearance of a level in addition to the (shifted) two-particle levels if there is a relatively narrow resonance in one channel. We have, for example, found additional levels related to the resonances $\rho$ \cite{Lang:2011mn}, $K^*(892)$ \cite{Prelovsek:2013ela}, $D^*_0(2400)$ \cite{Mohler:2012na},   and   the bound state $D_{s0}^*(2317)$ \cite{Mohler:2013rwa}. Additional levels related to $K_0^*(1430)$ \cite{Dudek:2014qha} and $X(3872)$ \cite{Prelovsek:2013cra}  have been found in the simulations of two coupled channels.   
Based on this experience, we expect an additional energy level if $Z_c$ is of similar origin, i.e. if it corresponds to a pole of the scattering matrix near physical axis.  
         
\section{Towards the lattice energy spectrum}
  
The energies  $E_n$ and the overlaps $Z_j^n\equiv \langle \Omega|{\cal O}_j|n\rangle$ of the eigenstates $n$ are extracted from the correlator matrix
\begin{equation}
\label{C}
C_{jk}(t)= \langle \Omega|{\cal O}_j (t_{src}+t) {\cal O}_k^\dagger (t_{src})|\Omega \rangle=\sum_{n}Z_j^nZ_k^{n*}~e^{-E_n t}~.
\end{equation}
The physical system for given quantum numbers is created from the vacuum $|\Omega\rangle$ using creation operators ${\cal O}_k^\dagger$  at time $t_{src}$ and the system propagates for time $t$ before being annihilated at $t_{sink}=t_{src}+t$ by ${\cal O}_j$. The creation/annihilation operators are called interpolators.  Our correlation matrix is averaged over every second $t_{src}$. 
    
We employ 22 interpolators  $O^{M_1M_2}$ that couple well to the two-meson states and the choice is expected to be complete enough to render all two-meson states listed in  (\ref{MM}). In addition, we implement  4 diquark-antidiquark interpolators ${\cal O}^{4q}$ with structure  $[\bar c\bar d]_{3_c}[cu]_{\bar 3_c}$ which is expected to couple well to possible $Z_c^+$ if it has a sizable Fock component of this kind. We point out that  ${\cal O}^{4q}\simeq [\bar c\bar d]_{3_c}[cu]_{\bar 3_c}$ couples also to two-meson states via Fierz rearrangement. Representative examples of employed interpolators are
 \begin{align} 
\label{O_main}
 &{\cal O}_1^{\psi(0)\pi(0)}=\bar c \gamma_i c(0)~\bar d\gamma_5 u(0)\,,\\ 
 &{\cal O}^{\psi(1)\pi(-1)}=\!\!\!\! \sum_{e_k=\pm e_{x,y,z}}\!\!\!~\bar c \gamma_i c(e_k)~\bar d\gamma_5 u(-e_k)\,, \nonumber\\
 & O^{\psi(2)\pi(-2)}\!\!=\!\!\!\! \sum_{|u_k|^2=2}~\bar c \gamma_i c(u_k)~\bar d\gamma_5 u(-u_k)\,, \nonumber\\
  &{\cal O}^{\eta_c(0)\rho(0)}=\bar c \gamma_5 c(0)~\bar d\gamma_i u(0)\,, \nonumber\\
  &{\cal O}_1^{D(0)D^*(0)}=\bar c \gamma_5 u(0)~\bar d\gamma_i c(0)   +  \{\gamma_5 \leftrightarrow \gamma_i\}\,,\nonumber  \\  
  &{\cal O}^{D^*(0)D^*(0)}=\epsilon_{ijk}~\bar c \gamma_j u(0)~\bar d\gamma_k c(0)\,, \nonumber  \\ 
&{\cal O}^{4q}_1\propto \epsilon_{abc} \epsilon_{ab'c'}(\bar c_{b}C \gamma_5\bar d_{c}~   c_{b'}\gamma_{i} C u_{c'}- \bar c_{b}C \gamma_i\bar d_{c}~  c_{b'}\gamma_{5} C u_{c'})\nonumber \,,\\ 
&{\cal O}^{4q}_2\propto \epsilon_{abc} \epsilon_{ab'c'}(\bar c_{b}C \bar d_{c}~   c_{b'}\gamma_{i} \gamma_5C u_{c'}- \bar c_{b}C \gamma_i\gamma_5\bar d_{c}~  c_{b'} C u_{c'})~\nonumber \,,
 \end{align}
while the full list of interpolators together with related details is provided in Appendix \ref{app_a}.
 
The momenta are projected separately for each meson   in $O^{M_1M_2}$ as 
$M(\mathbf{k})\simeq   \bar q_1\Gamma q_2(\mathbf{k})\equiv  \sum_{\mathbf{x}}e^{i2\pi \mathbf{k}\cdot\mathbf{x}/L}\bar q_1(\mathbf{x},t)\Gamma q_2(\mathbf{x},t)$. All quark fields   are smeared   according to the distillation method \cite{Peardon:2009gh,Mohler:2012na}.
      
The Wick contractions for the matrix of correlators (\ref{C}) with $I\!=\!1$ involve only diagrams where the light quarks $\bar d$ and $u$ propagate from source to sink. Concerning charm quarks, there are diagrams where they propagate from source to sink and diagrams where charm quarks annihilate  (Fig. \ref{fig:contractions} in Appendix \ref{app_b}).  The second class  represents mixing  with channels that contain no charm quarks, their effect is  suppressed due to the Okubo-Zweig-Iizuka rule, and the experiments do not observe these decay channels in the region of interest. The results in the present work are therefore based on the contractions in Fig. \ref{fig:contractions}a, where charm quarks propagate from source to sink. 

The energies $E_n$ and overlaps $Z_j^{n}$ are obtained from the
$22\times 22$ correlator matrix  (\ref{C}) using the generalized
eigenvalue method \cite{Michael:1985ne,Luscher:1985dn,Luscher:1990ck,Blossier:2009kd}
\begin{align} 
 C(t)u^{(n)}(t)&=\lambda^{(n)}(t)C(t_0)u^{(n)}(t)\,.
\end{align}
The energies $E_n$ are extracted from the asymptotically exponential
behaviour of the eigenvalues: $\lambda^{(n)}(t)\propto  e^{-E_n t}$ at
large $t$.  We use correlated two-exponential fits to $\lambda^{(n)}(t)$ where consistent results are found for $t_0\!=\!2,3$ and we present them for $t_0\!=\! 2$.    The errors-bars correspond to statistical errors obtained using the single-elimination jack-knife. 
Overlap factors follow from
\begin{equation}
\label{Z}
Z_j^{(n)}(t)=\mathrm{e}^{E_n t/2}  \frac{|C_{jk}(t) u_k^{(n)}(t)|}{|C(t)^\frac{1}{2} u^{(n)}(t)| } \;,
\end{equation}
fitted to a constant in $6\le t \le 11$.   

The treatment of the charm quarks requires special care due to discretization errors. We employ the Fermilab method \cite{ElKhadra:1996mp,Oktay:2008ex}, where discretization uncertainties are suppressed in the difference $E_n-m_{s.a.}$ with the spin-average mass $m_{s.a.}\!\equiv\! \tfrac{1}{4}(m_{\eta_c}\!+\!3m_{J/\psi})$.  The same method and tuning of the charm quark mass $m_c$ lead to a good agreement with experiment for conventional charmonium \cite{Mohler:2012na}, masses and widths of $D$ mesons \cite{Mohler:2012na}, and the $D_s$ spectrum \cite{Mohler:2013rwa,Lang:2014yfa} on this ensemble. In view of this, we will compare $E^{lat}_n-m_{s.a.}^{lat}+m_{s.a.}^{exp}$ to experiment where  $am_{\eta_c}^{lat}\!=\!1.47392(31)$ and $am_{J/\psi}^{lat}\!=\!1.54171(43)$. 
 
 \begin{figure*}[htb]
\begin{center}
\includegraphics[width=0.95\textwidth,clip]{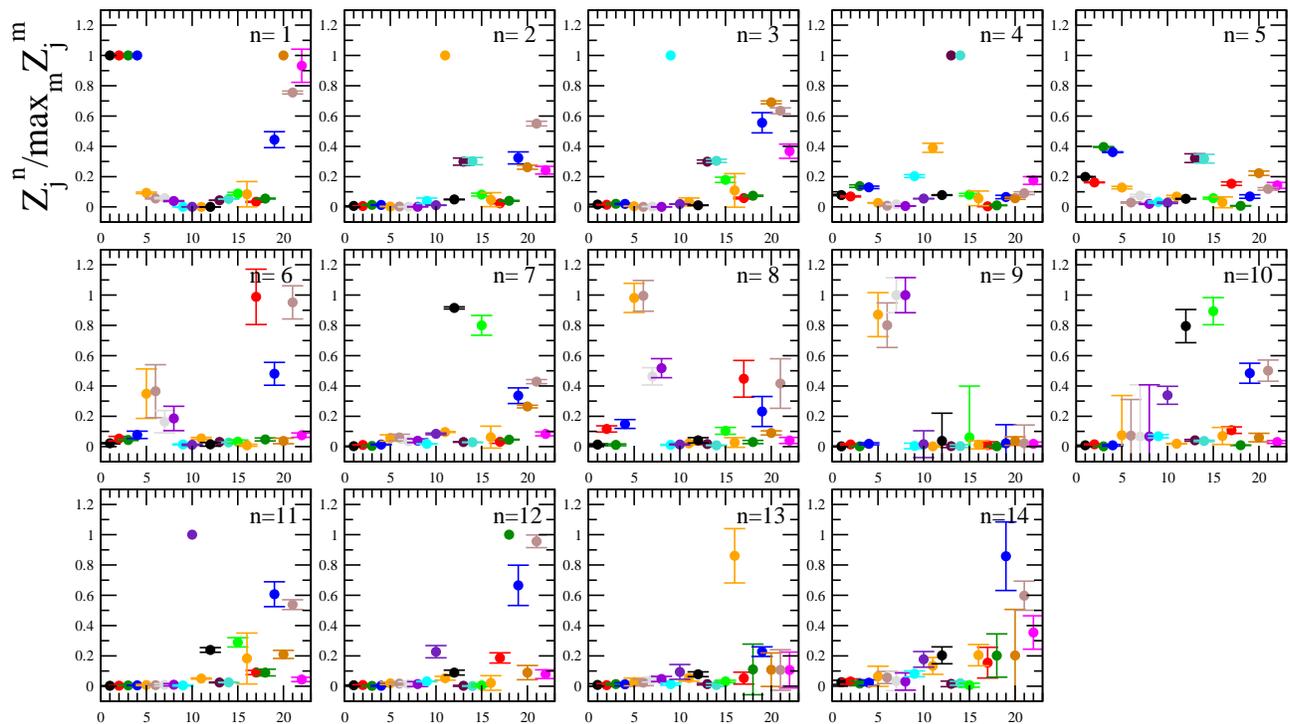}
 \end{center}
\caption{  The overlaps $Z_j^n=\langle \Omega|{\cal O}_j|n\rangle$  show the matrix elements of interpolators ${\cal O}_j$ between the vacuum $\langle \Omega|$ and the physical eigenstate $|n\rangle$ on the lattice. We present the overlap ratios $Z_j^n/\mathrm{max}_m Z_j^m$ where the denominator  is  the maximal  $|Z_j^m|$ at given operator number $j$. These ratios are  independent on the normalization of the interpolators $O_j$.  Levels $n=1,..,14$  are ordered  from lowest to highest $E_n$ in Fig. \ref{fig:eff}b. The horizontal axis denotes $j=1,..,22$ corresponding to complete basis of interpolators  ${\cal O}_j$  (\ref{O}). }\label{fig:Zratio} 
\end{figure*} 

\section{Results}\label{sec_results}
 
The central result of our simulation is the discrete spectrum in  Fig. \ref{fig:eff}b,  while experimental candidates in the same channel are collected in  Fig. \ref{fig:eff}a. In the energy region below $E\leq 4.3~$GeV one expects thirteen discrete two-particle states (\ref{MM})  near the horizontal lines, which continue in Fig. \ref{fig:eff}a to show their relation to the continuum of scattering states in experiment.

We interpret the lowest thirteen levels (indicated by black circles) as interacting two-particle states for the following reasons:
\begin{itemize} 
\item The levels  appear  near the non-interacting energies (\ref{E}) of the two-particle states (\ref{MM}). 
\item Each of these levels $n$ has the largest overlap  with the corresponding $O^{M_1M_2}$. This is shown in Fig. \ref{fig:Zratio} where 
overlaps $\langle \Omega|{\cal O}_j|n\rangle$ are provided in the  form of 
ratios $Z^n_j/ \mathrm{max}_m Z^m_j$  that do not depend on the normalization of ${\cal O}_j$ (\ref{O}).  The same conclusion applies for   $Z_j^n$ in Fig. \ref{fig:Zfactor}   which depend on the normalization  of ${\cal O}_j$ (\ref{O}).
\item When one of $O^{M_1M_2}$ is omitted  from the  correlator matrix, the corresponding two-particle level disappears from the spectrum or becomes very noisy.  This also indicates that the two-particle states are either decoupled or cannot be reliably extracted for the basis without the corresponding interpolators.
\item The thirteen lowest levels remain unaffected after ${\cal O}_{1-4}^{4q}$ are excluded from the interpolator basis. This can be seen by comparing the energy spectra in Figs. \ref{fig:eff}b and \ref{fig:eff}c, that show the result from the complete $22\times 22$ and the truncated $18\times 18$ correlation matrices. We verified that the thirteen lowest levels have very similar $Z_j^n$ for both choices of basis. 
\end{itemize}

The energy level $n=14$ at $E\simeq 4.39$ GeV in Fig. \ref{fig:eff}b (shown in green) seems like a sought state that appears in addition to thirteen expected two-meson states (\ref{MM}). This eigenstate  also has largest overlap with the tetraquark interpolating fields ${\cal O}^{4q}$ in Figs. \ref{fig:Zratio} and \ref{fig:Zfactor}. It might seem tempting to relate this level to a possible $Z_c^+$ candidate. However, the level $n=14$ lies close to  the expected two-meson states above $4.3~$GeV that we have omitted in the list  (\ref{MM}) since our aim was to search for candidates below $4.2~$GeV. Although the eigenstate $n=14$ might have an interesting structure, we cannot attribute this level to $Z_c^+$ candidate as we cannot rule out that it corresponds to one of omitted two-meson states above $4.3~$GeV. 

The main conclusion of our simulation is that we do not find any additional state below $4.2~$GeV   that could be related to an exotic candidate. We only find the expected two-meson states (\ref{MM}). 

 It is indeed surprising that with a basis (\ref{O}), which contains a great variety of interpolating fields with the quantum numbers of interest ($I^G(J^{PC})=1^+(1^{+-})$),  one does not, for example, induce  $Z_c(3900)/Z_c^+(3885)$ that has been confirmed by several experiments \cite{Ablikim:2013mio,Liu:2013dau,Xiao:2013iha,Ablikim:2013xfr}. Note that our list of creation/annihilation operators (\ref{O}) contains also a number of field structures $J/\Psi\pi$ and $D\bar D^*$ which correspond to channels where these resonances have been found in experiments  
  
   We list several possible reasons for the  absence of an energy levels related to the exotic $Z_c^+$ candidate in our simulation:
 \begin{itemize}
  \item The resonance $Z_c^+(3900)$ was found in $J/\psi\,\pi$ invariant mass only through $e^+e^- \to Y(4260)\to  (J/\psi\, \pi^+)\pi^- $ \cite{Ablikim:2013mio,Liu:2013dau,Xiao:2013iha}. No resonant structure in $J/\psi\, \pi^+$ invariant mass was seen in  $\bar B^0\to (J/\psi\, \pi^+) K^-$ by BELLE \cite{Chilikin:2014bkk}, in $\bar B^0 \to (J/\psi \pi^+) \pi^-$  by LHCb  \cite{Aaij:2014siy} or in  $\gamma p\to (J/\psi \,\pi^+) n$ by COMPASS    \cite{Adolph:2014hba}.     This might indicate that  the peak seen in   $e^+e^- \to Y(4260)\to  (J/\psi\, \pi^+)\pi^- $  might not be of dynamical origin. 
  \item  Along similar lines, several theoretical approaches  render peaks in $J/\psi \pi$ invariant mass for $e^+e^- \to Y(4260)\to \pi^- (J/\psi \pi^+) $ without invoking an exotic state. This is for example reproduced   as a coupled-channel (sometimes called cusp) effect  in  \cite{Chen:2013coa,Swanson:2014tra}.  
    \item The $J^P$  of  $Z_c(3900),~ Z_c(4020)$ and $Z_c(4025)$ is currently unknown from  experiment.  This might be the reason for their absence   in a simulation of  $1^+$ channel. We view this possibility as unlikely, since most previous studies favour $1^+$ for these states \cite{Brambilla:2014aaa}. 
  \item Even if the $Z_c^+$ resonant structure seen in experiment is due to a relatively narrow $\bar cc\bar du$ state, there might be several reasons that an additional state is absent in our simulation. It is possible that $Z_c$  exists only at physical $m_{u/d}$  and is absent at  unphysical $m_{u/d}$ in our simulation. Furthermore,   our set of eighteen interpolators ${\cal O}^{MM}$ may not be complete enough to render a $Z_c^+$ candidate in addition to thirteen two-meson states, even if $Z_c^+$ existed at $m_\pi=266~$MeV. 
  \item If significant S-wave D-wave mixing is vital for creating the observed experimental spectrum our setup might be unsuitable and we would probably miss an energy level emerging from this mixing.
  \item  Based on the experience discussed in Section II we would expect an 
additional energy level if the $Z_c^+$ state was a resonance
associated to pole near the real axis in the unphysical Riemann sheet.
The absence of an additional energy level 
could also indicate a different origin of the experimental peak like, 
e.g., a coupled-channel threshold effect.  
\end{itemize}

 \section{Cautionary remarks} \label{sec_cautionary}
 
In this search for exotics from first-principle QCD we have learned some lessons and we thus collect here
some cautionary remarks. 
 
 \subsection{ Consideration of  the ground state }
 
In lattice QCD studies and QCD sum rule studies one is sometimes tempted to draw conclusions on the exotic states by looking at the ground state obtained from correlators of type $\langle \Omega |{\cal O}^{4q\dagger}{\cal O}^{4q}|\Omega\rangle$ or $\langle \Omega |{\cal O}^{M_1M_2\dagger}{\cal O}^{M_1M_2}|\Omega\rangle$. This may be misleading.
 
\begin{figure*}[htb]
\begin{center}
\includegraphics[width=0.79\textwidth,clip]{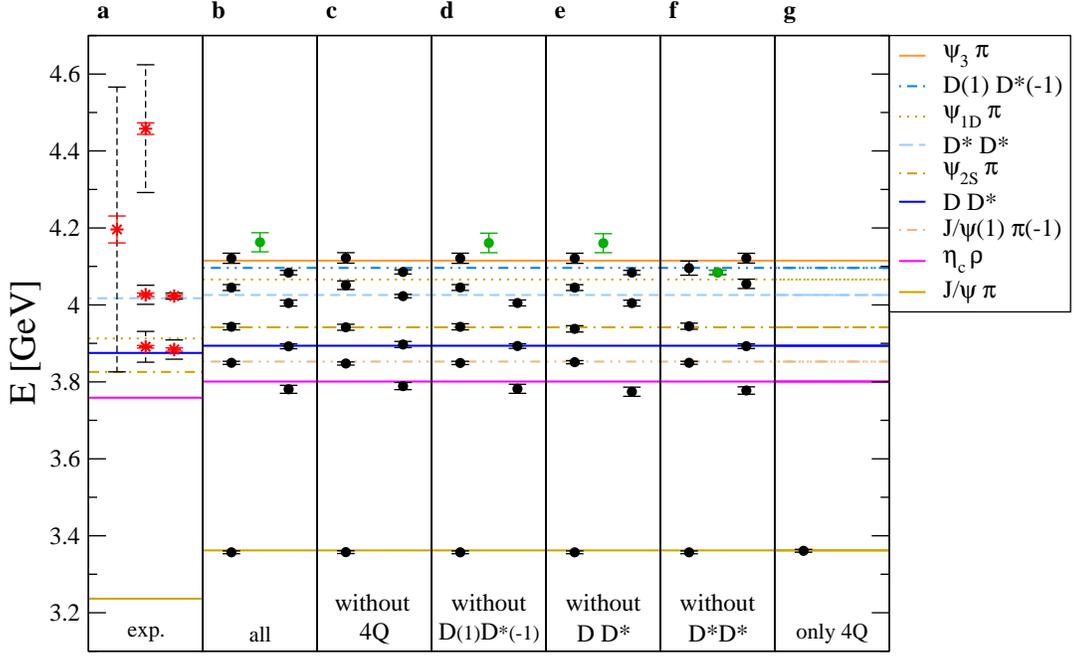} 
\end{center}
\caption{The spectrum for   $ I^G(J^{PC})=1^+(1^{+-})$ for the case of the reduced interpolator basis (\ref{O_cautionary}) with the purpose of illustrating caveats in Section \ref{sec_reduced}.  (b)  Lattice spectrum based on the $18\times 18$ correlator matrix (\ref{O_cautionary}); (c)   spectrum based on a $14\times 14$ correlator matrix without diquark-antidiquark interpolating fields ${\cal O}^{4q}_{1-4}$; spectra (d-f) are based on truncated correlator matrices as described in the figure; spectrum (g) is based on ${\cal O}^{4q}_{1-4}$ only.  The horizontal lines represent energies of the non-interacting two-particle states.  Statistical errors on the lattice spectrum are shown.  }\label{fig:eff_cautionary}
\end{figure*}

 The ground state for the quantum numbers in our study is $J/\psi\, \pi$ and we observe it for most interpolators at large $t$ after the exponentials due to higher states have died out. This applies also for  $\langle {\cal O}^{4q}(t)|{\cal O}^{4q}(0)\rangle\propto e^{-(m_{J/\psi}+m_\pi)t}$ at large $t$, as shown in Fig. \ref{fig:eff_cautionary}g. Looking at the ground state of the diquark-antidiquark correlators alone, one cannot reach conclusions regarding $Z_c^+$. This holds also for the  ground states from ${\cal O}^{D D^*}$ (used in \cite{Chen:2014afa})  or ${\cal O}^{D^* D^*}$ correlators alone. The coupling to $J/\psi\, \pi$,  $\eta_c\rho$  (and possibly some others) has to be taken into account, as shown by our study. These cautionary remarks also apply to QCD sum-rule studies that are based on correlators. 

 \subsection{Reduced interpolator basis}\label{sec_reduced}
  
  Here we show an example why the $Z_c^+$ candidate  is not reliable as long as not all  two-meson states with lower energy values and at least one nearby state with higher energy have been established.  We illustrate that by a simulation \cite{Prelovsek:2014v1} that aims at extracting nine two-meson states $J/\psi(0)\pi(0),\,\eta_c(0)\rho(0),~J/\psi(1)\pi(-1),~D(0)\bar D^*(0),$ $\psi_{2S}(0)\pi(0),$$~D^*(0)\bar D^*(0),\psi_{1D}(0)\pi(0),~D(1)\bar D^*(-1),$ $\psi_3(0)\pi(0)$ 
 using a correspondingly chosen interpolator basis (numbered according to (\ref{O}))
 \begin{equation}
 \label{O_cautionary}
 {\cal O}_{1-9, ~11,~13-15,~ 17,~ 19-22} 
 \end{equation}
which should suffice for the extraction of the mentioned nine two-meson states and possibly an additional exotic candidate.
 
 The spectrum in Fig. \ref{fig:eff_cautionary}b and overlaps in Fig. \ref{fig:Zfactor_cautionary} show that the lowest nine states (indicated by black circles) are two-meson states.  When either one of $O^{D(0)D^*(0)}$, $O^{D(1)D^*(-1)}$ or $O^{D^*(0)D^*(0)}$ is omitted from the  correlator matrix, the corresponding two-particle level disappears from the spectrum (Figs. \ref{fig:eff_cautionary}d-\ref{fig:eff_cautionary}f). 
 
 The state $n=10$ at $E\simeq 4.16~$GeV (shown in green) is an extra state and it has large overlap with diquark-antidiquark interpolators in Fig. \ref{fig:eff_cautionary}.  Figure  \ref{fig:Zfactor_cautionary} shows that this state disappears from the spectrum if ${\cal O}^{4q}$ are omitted from the basis (\ref{O_cautionary}), so the ${\cal O}^{4q}$ Fock component seems to be crucial for its existence. We also verified that the energy of the extra state is rather stable under different choices of  ${\cal O}^{MM}$ among (\ref{O_cautionary}), as long as the ${\cal O}^{4q}$ are kept in the basis. This led to the premature conclusion \cite{Prelovsek:2014v1}  that an extra level $n=10$ could be related to a $Z_c^+$ candidate. 
 
 The results  in Section \ref{sec_results} from the complete interpolator basis (\ref{O}), which incorporate also two-meson states in the region between $4.2$ and $4.3$ GeV, do not show an additional state near $4.16~$GeV.  Furthermore the complete basis renders an additional state at higher energy $E\simeq 4.39~$GeV, therefore the state from Fig. \ref{fig:eff_cautionary} at  $E\simeq 4.16~$GeV is not a reliable candidate. 
 
 In fact, the appearance of an additional state close to the energy region of the first omitted two-meson states in Figs. \ref{fig:eff} and \ref{fig:eff_cautionary} seems to indicate that such an additional state may be related to (a linear combination of)  omitted two-meson states via  ${\cal O}^{4q}$.  This is not surprising as $[\bar c\bar d]_{3_c}[cu]_{\bar 3_c}$  contains a linear combination of various $M_1(k)M_2(-k)$ after the Fierz rearrangement.  In light of this, it is puzzling  that the ${\cal O}^{4q}$ interpolators do not render lower lying two-meson states $M_1(k)M_2(-k)$  if the corresponding ${\cal O}^{M_1(k)M_2(-k)}$ are omitted from the basis. This is shown in Figs. \ref{fig:eff_cautionary}d-\ref{fig:eff_cautionary}f for the cases if either one of $D(0)D^*(0)$, $D(1)D^*(-1)$ or $D^*(0)D^*(0)$ is omitted.  
 
 From this we conclude  that an exotic candidate may be reliable only when all two-meson states below it and at least one slightly above it have also been established in the lattice simulation. 
 
 \subsection{Aiming at larger volumes}\label{sec_V}
 
 A lattice study  at larger lattice size $L$ will involve even more states $M_1(k)M_2(-k)$ below a certain energy according to (\ref{E}). The present number of thirteen two-mesons states below $4.3~$GeV at $L\simeq 2~$fm would significantly increase for a simulation at $L=3~$fm. This would require additional scattering operators thus increasing the correlator matrix size.  Due to the larger physical volume more eigenvectors would be required in the distillation method.

 \section{Conclusions}   
 
We presented a lattice QCD simulation for the $\bar c c \bar d u$ channel with  $J^{PC}=1^{+-}$ where exotic charmonia have been found in recent experiments; the pion mass in our study is $266$ MeV. 
In our set of 22 interpolating operators we allow  for all possible meson-meson operators in the energy region between the $J/\psi \pi$ threshold and 4.3 GeV and we also introduce four diquark-antidiquark operators.
In the scanned energy region we find all expected meson-meson signals (mostly close to the non-interacting levels) but
no convincing signal for an extra $Z_c^+$ state. Possible physics and methodology-related reasons for the absence of the exotic candidate in our simulation are mentioned. We also discuss in detail possible traps leading to premature identifications. We conclude that at least with the diquark-antidiquark and meson-meson operators used here our ab-initio study shows no
exotic state below 4.2 GeV.

\begin{appendix}
\section{Interpolators}\label{app_a}
 
 We implement altogether 22 interpolators with $I^G=1^+$,   $J^{PC}=1^{+-}$  and total momentum zero (using the irreducible representation $T_1^{+-}$ of the lattice symmetry group $O_h$). The first 18 interpolators ${\cal O}^{M_1M_2}$ are expected to couple well to the two-meson states (\ref{MM}) , while the last four are diquark-antidiquark interpolators $O^{4q}$ with structure $[\bar c\bar d]_{3_c}[cu]_{\bar 3_c}$
\begin{align} 
\label{O}
{\cal O}_1&= {\cal O}_1^{\psi(0)\pi(0)}=\bar c \gamma_i c(0)~\bar d\gamma_5 u(0)\,,\\
{\cal O}_2&= {\cal O}_2^{\psi(0)\pi(0)}=\bar c \gamma_i \gamma_t c(0)~\bar d\gamma_5  u(0)\,,\nonumber \\ 
{\cal O}_3&=  {\cal O}_3^{\psi(0)\pi(0)}=  \bar c \overleftarrow{\nabla}_j\gamma_i \overrightarrow{\nabla}_j c(0)~\bar d\gamma_5 u(0)\,, \nonumber\\
{\cal O}_4&=  {\cal O}_4^{\psi(0)\pi(0)}=  \bar c \overleftarrow{\nabla}_j\gamma_i\gamma_t \overrightarrow{\nabla}_j c(0)~\bar d\gamma_5 u(0) \,, \nonumber\\
{\cal O}_5&=  {\cal O}_5^{\psi(0)\pi(0)}=  |\epsilon_{ijk}| |\epsilon_{klm}|~\bar c \gamma_j \overleftarrow{\nabla}_l  \overrightarrow{\nabla}_m c(0)~\bar d\gamma_5 u(0) \,, \nonumber\\
{\cal O}_6&= {\cal O}_6^{\psi(0)\pi(0)} =  |\epsilon_{ijk}| |\epsilon_{klm}| ~\bar c \gamma_t \gamma_j  \overleftarrow{\nabla}_l  \overrightarrow{\nabla}_m c(0)~\bar d\gamma_5 u(0)\,, \nonumber
\end{align}
\begin{align}
{\cal O}_7&=  {\cal O}_7^{\psi(0)\pi(0)}=R_{ijk} Q_{klm}~ \bar c  \gamma_j  \overleftarrow{\nabla}_l\overrightarrow{\nabla}_m c~\bar d\gamma_5 u(0)  \,, \nonumber\\
{\cal O}_8&= {\cal O}_8^{\psi(0)\pi(0)}=R_{ijk} Q_{klm}~ \bar c  \gamma_t \gamma_j  \overleftarrow{\nabla}_l\overrightarrow{\nabla}_m c~\bar d\gamma_5 u(0)\,,  \nonumber \\
{\cal O}_9&=  {\cal O}^{\psi(1)\pi(-1)}=\!\!\!\! \sum_{e_k=\pm e_{x,y,z}}\!\!\!~\bar c \gamma_i c(e_k)~\bar d\gamma_5 u(-e_k)\,, \nonumber\\
{\cal O}_{10}&= O^{\psi(2)\pi(-2)}\!\!=\!\!\!\! \sum_{|u_k|^2=2}~\bar c \gamma_i c(u_k)~\bar d\gamma_5 u(-u_k)\,, \nonumber\\
{\cal O}_{11}&= {\cal O}^{\eta_c(0)\rho(0)}=\bar c \gamma_5 c(0)~\bar d\gamma_i u(0)\,, \nonumber\\
{\cal O}_{12}&= O^{\eta_c(1)\rho(-1)}=\!\!\!\!\sum_{e_k=\pm e_{x,y,z}}\!\!\!\bar c \gamma_5 c(e_k)~\bar d\gamma_i u(-e_k)\,, \nonumber\\
{\cal O}_{13}&= {\cal O}_1^{D(0)D^*(0)}=\bar c \gamma_5 u(0)~\bar d\gamma_i c(0)   +  \{\gamma_5 \leftrightarrow \gamma_i\}\,,\nonumber  \\
{\cal O}_{14}&= {\cal O}_2^{D(0)D^*(0)}=\bar c \gamma_5 \gamma_t u(0)~\bar d\gamma_i \gamma_t c(0) +   \{\gamma_5 \leftrightarrow \gamma_i\} \,,\nonumber\\
{\cal O}_{15}&=  {\cal O}^{D(1)D^*(-1)}\!\!=\!\!\!\!\!\!\!\!\!\sum_{e_k=\pm e_{x,y,z}}\!\!\!\!\!\!\bar c \gamma_5 u(e_k)~\bar d\gamma_i c(-e_k) +  \{\gamma_5 \leftrightarrow \gamma_i\}   \nonumber\,,\\
{\cal O}_{16}&= O^{D(2)D^*(-2)}\!\!=\!\!\!\! \sum_{|u_k|^2=2}\!\!\bar c \gamma_5 u(u_k)~\bar d\gamma_i c(-u_k) +  \{\gamma_5 \leftrightarrow \gamma_i\}   \nonumber\,,\\
{\cal O}_{17}&= {\cal O}^{D^*(0)D^*(0)}=\epsilon_{ijl}~\bar c \gamma_j u(0)~\bar d\gamma_l c(0)\,,   \nonumber\\
{\cal O}_{18}&= O^{D^*(1)D^*(-1)}=\!\!\!\!\sum_{e_k=\pm e_{x,y,z}}\!\!\!\epsilon_{ijl}~\bar c \gamma_j u(e_k)~\bar d\gamma_l c(-e_k)   \nonumber\\
{\cal O}_{19}&= {\cal O}^{4q}_1= N_L^3~\epsilon_{abc} \epsilon_{ab'c'}(\bar c_{b}C \gamma_5\bar d_{c}~   c_{b'}\gamma_{i} C u_{c'}\nonumber\\
  &\qquad\qquad\qquad\qquad\quad- \bar c_{b}C \gamma_i\bar d_{c}~  c_{b'}\gamma_{5} C u_{c'})\nonumber \,,\\ 
{\cal O}_{20}&= {\cal O}^{4q}_2= N_L^3~\epsilon_{abc} \epsilon_{ab'c'}(\bar c_{b}C \bar d_{c}~   c_{b'}\gamma_{i} \gamma_5C u_{c'}\nonumber\\
&\qquad\qquad\qquad\qquad\quad- \bar c_{b}C \gamma_i\gamma_5\bar d_{c}~  c_{b'} C u_{c'})\nonumber\,,\\ 
{\cal O}_{21}&= {\cal O}^{4q}_3={\cal O}^{4q}_1(N_v\!=\!32)\nonumber\,,\\
{\cal O}_{22}&= {\cal O}^{4q}_4={\cal O}^{4q}_2(N_v\!=\!32)\,.\nonumber 
\end{align}
 We implicitly sum over index pairs. In the definitions, $u_k$ indicates a sum over 12 directions $\pm e_i\pm e_j$ with $|u_k|^2=2$, 
 $i$ denotes the polarisation and the correlation matrix is averaged over $i=x,y,z$. 
The $Q_{ilm}$ are taken from \cite{Mohler:2012na} and $R_{ilm}$ from \cite{Dudek:2007wv}. 
 
The momenta are projected separately for each meson  $M_1(\mathbf{k})M_2(-\mathbf{k})$ in $O^{M_1M_2}$ 
 \begin{equation}
M(\mathbf{k}):\ \  \bar q_1\Gamma q_2(\mathbf{k})\equiv  \sum_{\mathbf{x}}e^{i2\pi \mathbf{k}\mathbf{x}/L}q_1(\mathbf{x},t)\Gamma q_2(\mathbf{x},t)~\quad \mathbf{k}\in \mathbb{N}^3\,.
 \end{equation}
The momentum in ${\cal O}^{4q}$ is projected to zero,  
  \begin{equation}
\label{4q}
O^{4q}=N_L^3\epsilon_{abc} \epsilon_{ab'c'}\sum_{\mathbf{x}} \bar c_{b}(\mathbf{x},t)\Gamma_1 \bar d_{c}(\mathbf{x},t)~c_{b'}(\mathbf{x},t)\Gamma_{2} u_{c'}(\mathbf{x},t)\,,
\end{equation}  
and a factor $N_L^3$ is included to achieve similar normalization as for  $O^{M_1M_2}$.   The   $O^{4q}$  are implemented   as  
\begin{align}
\tilde O^{4q}\!&=\!N_L^3~\epsilon_{abc} \epsilon_{ab'c'}\\
&\sum_{\mathbf{x}_1} \bar c_{b}(\mathbf{x}_1,t)\Gamma_1 \bar d_{c}(\mathbf{x}_1,t)~\sum_{\mathbf{x}_2}c_{b'}(\mathbf{x}_2,t)\Gamma_{2} u_{c'}(\mathbf{x}_2,t)\nonumber
\end{align}
which reduces to $O^{4q}$ after the average over gauge configurations, where the gauge is not fixed. We verified  explicitly  that $\langle \tilde O^{4q}|\tilde O^{4q\dagger}\rangle\simeq \langle  O^{4q}|O^{4q\dagger}\rangle$. 

All quark fields  in (\ref{O}) are smeared  $q\equiv \sum_{k=1}^{N_v}v^{(k)}v^{(k)\dagger}q_{point}$ according to the distillation method \cite{Peardon:2009gh,Mohler:2012na}. We employ $N_v\!=\!64$ Laplacian eigenvectors for all interpolators with exception of ${\cal O}^{4q}_{3,4}$ where the smearing with $N_v\!=\!32$ is used. 

Eight interpolators ${\cal O}^{\psi(0)\pi(0)}_{1,..,8}$   are implemented in order to allow the reliable extraction of  two-meson states $\psi(0)\pi(0)$ with $\psi\!=\!J/\psi,\, \psi(2S),\,\psi(3770),\, \psi_3$. We verified for conventional charmonium that  eight $\bar cc$ structures in ${\cal O}^{\psi\pi}_{1,..,8}$ lead to a reliable signal for these four $\psi$'s after the diagonalization of  the $8\times 8$ correlation matrix. The first six $\bar cc$ structures were  used already in \cite{Mohler:2012na}, while  ${\cal O}_{7,8}^{\psi\pi}$ were added to enhance overlap with   $\psi_{3}$  \cite{Dudek:2007wv}. The $\psi_3$ denotes the charmonium with  $J^{PC}\!=\!3^{--}$ and appears in addition to $1^{--}$ states when  charmonia are simulated  using the irreducible representation $T_{1}^{--}$. This is a consequence of  the broken rotational invariance on a lattice, where the continuous symmetry group  is reduced to  $O_h$.  In order to study the $J^{PC}=1^{+-}$ channel in this work, we employ lattice interpolators that transform according to irreducible representation $T_1^{+-}$ of $O_h$. This irreducible representation contains  $J^{PC}=1^{+-}$ states, but  also the $\psi_{3}\, \pi$ state with $J^{PC}=3^{+-}$.  

We do not implement interpolators corresponding to $h_c(1)\pi(-1)$, since the simplest choice $\epsilon_{ijk}~[\bar c \gamma_j\gamma_t\gamma_5 c(e_k)~\bar d\gamma_5 u(-e_k)-\bar c \gamma_j\gamma_t\gamma_5 c(-e_k)~\bar d\gamma_5 u(e_k)] $ renders sizable coupling of $\bar c \gamma_y\gamma_t\gamma_5 c(e_x)$ also to lower-lying $J/\psi(e_x)$. The $h_c(1)\pi(-1)$ lies near $4.25~$GeV and omission of this two-meson state does not modify our physics conclusion concerning absence of $Z_c^+$ candidate below $4.2~$GeV.

\section{Wick contractions}\label{app_b}
 
\begin{figure*}[h!]
\begin{center}
\includegraphics[width=0.55\textwidth,clip]{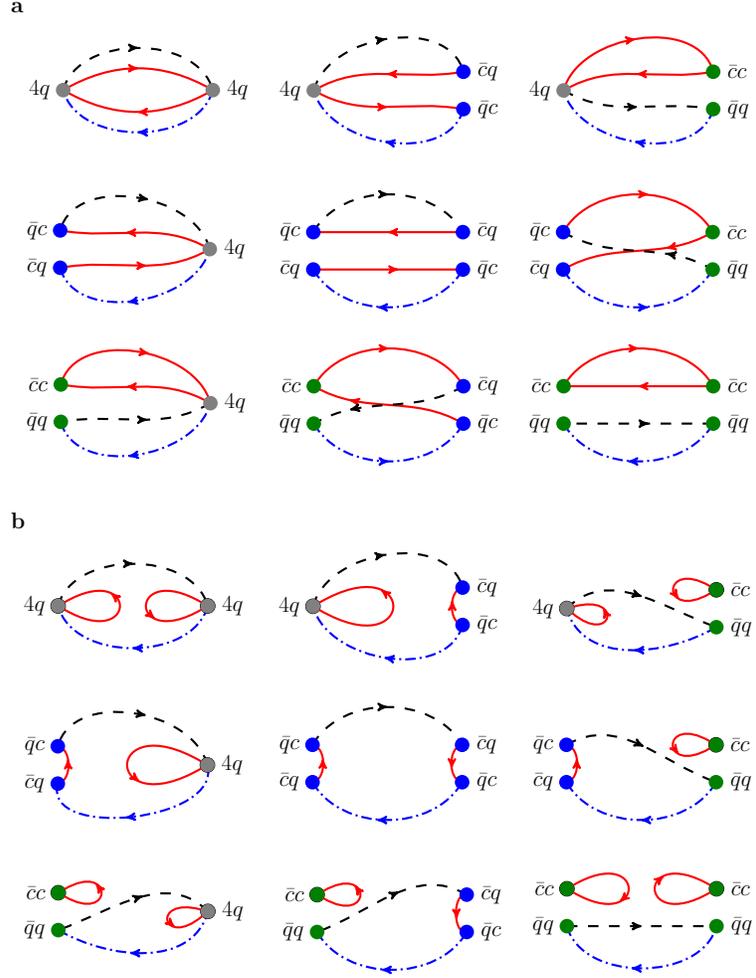}
 \end{center}
\caption{ Wick contractions that enter the correlation matrix $C_{jk}(t)$  for the interpolators (\ref{O}). A red solid line represents a $c$ quark, a black dashed line represents $u$, and the blue dash-doted line stands for $d$. (a) Nine diagrams where charm does not annihilate. (b) Nine diagrams where charm annihilates.}\label{fig:contractions}
\end{figure*}

 The Wick contractions that appear in the correlation matrix (\ref{C}) for interpolators (\ref{O}) are drawn in Fig. \ref{fig:contractions}. Our correlation matrix  is based on Wick contractions in Fig. \ref{fig:contractions}a, as explained in the main text. 
 
\section{Overlaps $\langle \Omega|{\cal O}_j|n\rangle$ for all eigenstates}\label{app_c}

Here we present the overlaps $\langle \Omega|{\cal O}_j|n\rangle$ of eigenstates to employed interpolators. These show which Fock components are important for various eigenstates. Note that the overlap factors $Z_j^n$ depend on the normalisation of ${\cal O}_j$ (\ref{O}), while the ratios presented in Fig. \ref{fig:Zratio} are independent of it. 

\begin{figure*}[htb]
\begin{center}
\includegraphics[width=0.95\textwidth,clip]{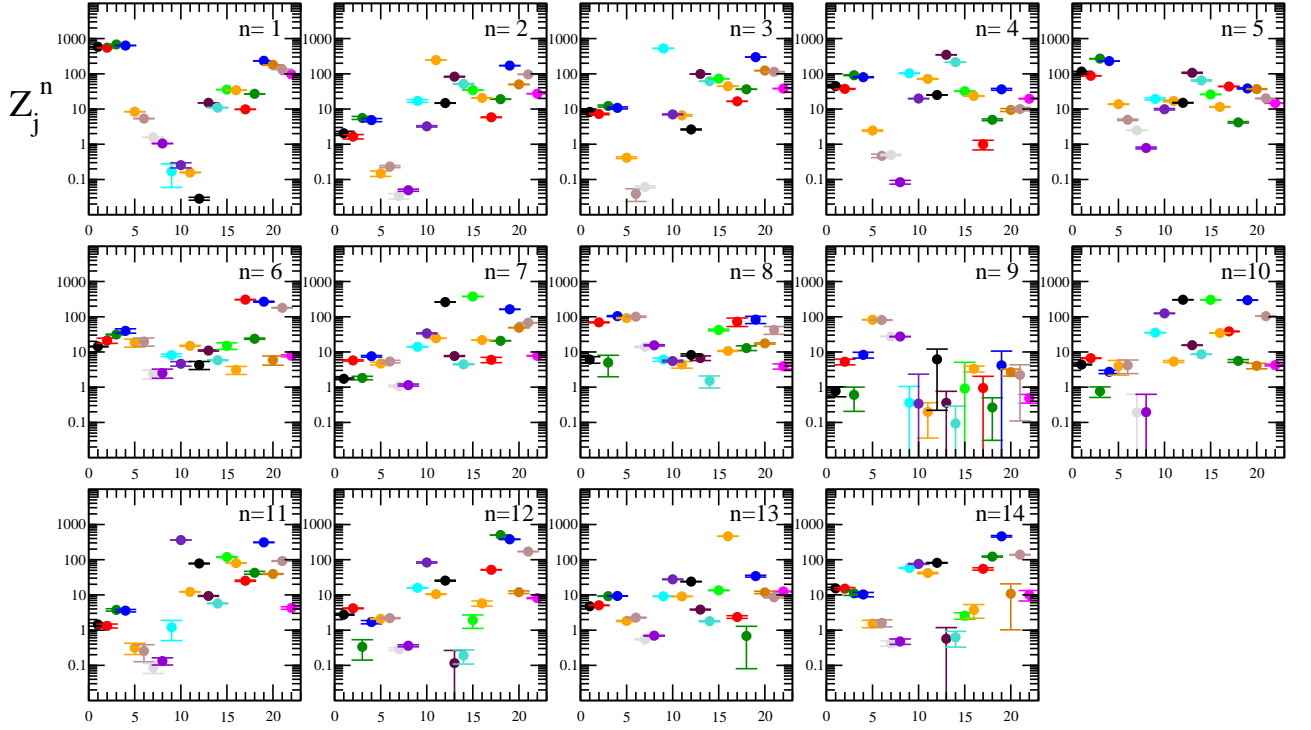}
 \end{center}
\caption{  The overlaps $Z_j^n=\langle \Omega|{\cal O}_j|n\rangle$  show the matrix elements of interpolators ${\cal O}_j$ between the vacuum $\langle \Omega|$ and the physical eigenstate $|n\rangle$ on the lattice.    Levels $n=1,..,14$  are ordered  from lowest to highest $E_n$ in Fig. \ref{fig:eff}b. The horizontal axis denotes $j=1,..,22$ corresponding to complete basis of interpolators  ${\cal O}_j$  (\ref{O}). }\label{fig:Zfactor} 
\end{figure*}

\begin{figure*}[h!]
\begin{center}
\includegraphics[width=0.95\textwidth,clip]{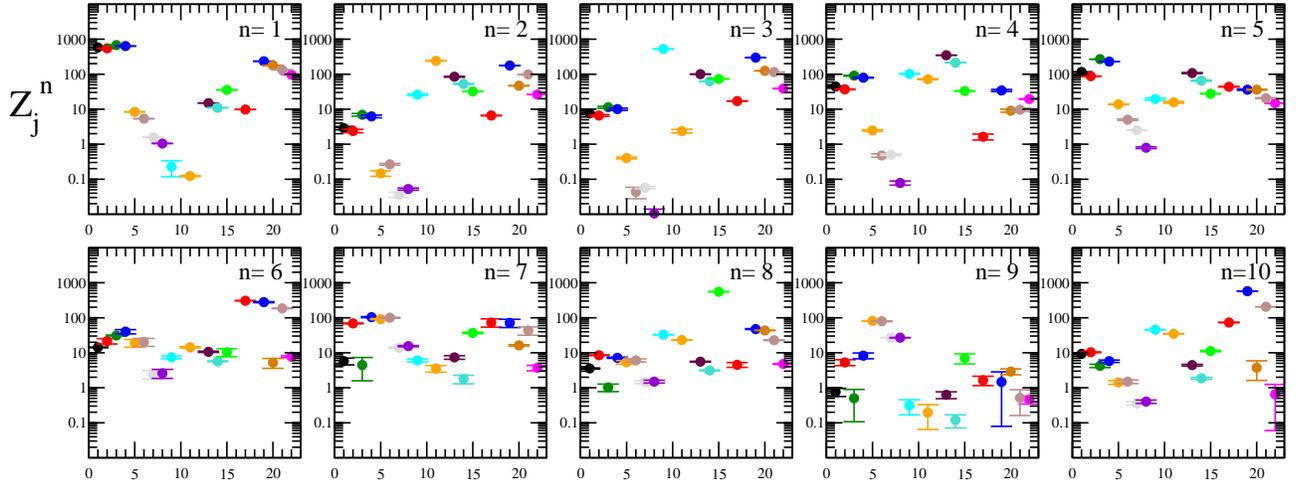}
 \end{center}
\caption{  The overlaps $Z_j^n=\langle \Omega|{\cal O}_j|n\rangle$  corresponding to $18\times 18$ correlation matrix for the reduced basis listed  in (\ref{O_cautionary}). The horizontal axis denotes $j$ corresponding to basis of interpolators  ${\cal O}_j$  (\ref{O}). }\label{fig:Zfactor_cautionary} 
\end{figure*}

The complete basis of $22$ interpolators (\ref{O}) leads to the overlaps in Fig. \ref{fig:Zfactor}. The result from the reduced basis of $18$ interpolators (\ref{O_cautionary}) is presented in Fig. \ref{fig:Zfactor_cautionary} to illustrate the cautionary remarks discussed in Section \ref{sec_cautionary}.  
 

\end{appendix}
 
\acknowledgments
We thank Anna Hasenfratz for providing the gauge configurations. S.P. thanks Changzheng 
Yuan and An\v ze Zupanc for discussion, D.M. is grateful for discussions with Jim Simone, and C.B.L. for discussion with M. Padmanath.  The computations were done on the clusters at the Theoretical Physics department of Jozef Stefan Institute, at the University of Graz, NAWI Graz, and TRIUMF.  We acknowledge the support by the Slovenian Research Agency ARRS project N1-0020, by the Austrian Science Fund FWF project I1313-N27 and by DFG project SFB/TRR55. Fermilab is operated by Fermi Research Alliance, LLC under Contract No. De-AC02-07CH11359 with the United States Department of Energy.

\end{document}